\documentclass[twocolumn,showkeys,aps,amssymb,floatfix,prd]{revtex4}
\usepackage{bm,graphicx}

\newcommand\beq{\begin{equation}}
\newcommand\beqa{\begin{eqnarray}}
\newcommand\beqan{\begin{eqnarray*}}
\newcommand\eeq{\end{equation}}
\newcommand\eeqa{\end{eqnarray}}
\newcommand\eeqan{\end{eqnarray*}}

\newcommand\Mbh{M_\bullet}
\newcommand\Obh{\Omega_\bullet}
\newcommand\gravr{{\sf m}_\bullet}

\newcommand\vthE{\vartheta_E}
\newcommand\vthl{\vartheta_\ell}
\newcommand\vem{\varepsilon_m}
\newcommand\vel{\varepsilon_\ell}
\newcommand\vth{\vartheta}
\newcommand\vthbh{\vartheta_\bullet}
\newcommand\cb{{\cal B}}
\newcommand\order[2]{\mathcal{O}\left({#1}\right)^{#2}}
\newcommand\rr{\bar{r}}
\newcommand\mm{{\cal M}}
\newcommand\dmin{d_{\rm min}}
\newcommand\E[1]{\times10^{#1}}
\newcommand\bh{{\rm bh}}
\newcommand\dm{{\rm dm}}

\newcommand\reffig[1]{Figure~\ref{fig:#1}}

\begin{document}

\title{Formalism for testing theories of gravity using lensing by \\
compact objects. III: Braneworld gravity}

\author{Charles R.\ Keeton}
\affiliation{Department of Physics \& Astronomy, Rutgers University,
  136 Frelinghuysen Road, Piscataway, NJ 08854;\\
  {\tt keeton@physics.rutgers.edu}}

\author{A.\ O.\ Petters}
\affiliation{Departments of Mathematics and Physics,
  Duke University,\\
  Science Drive,\\ Durham, NC 27708-0320;\\
  {\tt petters@math.duke.edu}
\\
\\
{\rm Accepted in Phys Rev D}
}

\begin{abstract}
Braneworld gravity is a model that endows physical space with an
extra dimension.  In the type II Randall-Sundrum braneworld
gravity model, the extra dimension modifies the spacetime geometry
around black holes, and changes predictions for the formation and
survival of primordial black holes.  We develop a comprehensive
analytical formalism for far-field black hole lensing in this model,
using invariant quantities to compute all the geometric optics
lensing observables: bending angle, image position, magnification,
centroid, and time delay.  We then make the first analysis of wave
optics in braneworld lensing, working in the semi-classical limit.
Through quantitative examples we show that wave optics offers the
only realistic way to observe braneworld effects in black hole
lensing.  We point out that if primordial braneworld black holes
exist, have mass $\Mbh$, and contribute a fraction $f_\bh$ of the
dark matter, then roughly
$\sim 3\E{5} \times f_\bh (\Mbh/10^{-18}\,M_\odot)^{-1}$ of them
lie within our Solar System.  These objects, which we call
``attolenses,'' would produce interference fringes in the energy
spectra of gamma-ray bursts at energies
$E \sim 100\,(\Mbh/10^{-18}\,M_\odot)^{-1}$ MeV (which will soon
be accessible with the GLAST satellite).
Primordial braneworld black holes spread throughout the universe
could produce similar interference effects.  If they contribute a
fraction $\Obh$ of the total energy density, the probability that
gamma-ray bursts are ``attolensed'' is at least $\sim\!0.1\,\Obh$.
If observed, attolensing interference fringes would yield a simple
upper limit on $\Mbh$.  Detection of a primordial black hole with
$\Mbh \lesssim 10^{-19}\,M_\odot$ would challenge general
relativity and favor the braneworld model.  Further work on
lensing tests of braneworld gravity must proceed into the
physical optics regime, which awaits a description of the full
spacetime geometry around braneworld black holes.
\end{abstract}


\maketitle

\section{Introduction}
\label{sec:intro}

Gravitational lensing has emerged as a powerful and far-reaching
tool in astrophysics and cosmology \cite{SEF,PLW,Saas-Fee}.
In this series we are showing how lensing can also be employed 
to test theories of gravity.  Papers  I \cite{paperI} and
II \cite{paperII} developed an analytical formalism for identifying
the lensing signatures of gravity models that fall within the
post-post-Newtonian (PPN) framework, even probing out to third
order in such models.  These studies uncovered some surprising
universal relations among lensing observables that helped us make
specific predictions that are testable with current or near-future
instrumentation.  

In this paper we examine a gravity model that lies outside the
PPN framework, namely, type II Randall-Sundrum braneworld gravity
\cite{RS}.  According to this model, familiar 4-dimensional
spacetime is actually a submanifold (a ``brane'') in a
5-dimensional spacetime (the ``bulk''), with the extra dimension
characterized by a curvature radius $\ell$ which could be as
large as $\sim$0.2 mm \cite{measure-ell}.  One intriguing
prediction of the model is that braneworld black holes might be
produced at energies as low as $\sim$1 TeV, which could lead to
observable Hawking radiation in the forthcoming Large Hadron
Collider \cite{bw-lhc} or create specific signatures in cosmic
ray showers \cite{bw-showers}.

Another important prediction is that braneworld black holes
produced in the early universe might survive to the present
day.  Primordial black holes are predicted to have formed
from density fluctuations in the very early universe, with
a mass spectrum that increases rapidly towards low masses
\cite{PBHspec-1,PBHspec-2}.  In general relativity, black
holes smaller than $\sim\!10^{-19}\,M_\odot$ would have
evaporated by now through Hawking radiation \cite{GRevap}.  
Compared with their GR counterparts, however, braneworld
black holes evaporate more slowly \cite{primordial-1} and
may have accreted more efficiently in the early universe
\cite{primordial-2,primordial-3}.  Together, these effects
may allow primordial braneworld black holes as small as
$\Mbh \sim 1 \mbox{ kg} \sim 10^{-30}\,M_\odot$ to survive
to today \cite{primordial-3,MajMuk1}.  The implication is
that primordial braneworld black holes may contribute some
fraction of the unknown dark matter.  Gravitational lensing
offers a crucial test of this hypothesis if we can identify
appropriate lensing scenarios that carry a clear imprint of
braneworld gravity.  That is our goal.

An exact metric for the spacetime geometry induced by a
braneworld black hole is still unknown.  In the far-field or
weak-deflection regime, it is well established that braneworld
black holes are described by the
Garriga-Tanaka metric \cite{GT,GKR,SSM}.  In the near-field or
strong-deflection regime, various metrics are believed to
approximate the true spacetime geometry (e.g., \cite{GWBD}).
These different metrics have been used to explore gravitational
lensing by braneworld black holes (see the review by Majumdar \&
Mukherjee \cite{MajMuk1}).
For example, Kar \& Sinha \cite{KS} computed the light bending
angle for the Garriga-Tanaka and tidal Reissner-Nordstr\"om
metrics.
Majumdar \& Mukherjee \cite{MajMuk2} determined the light
bending angle, image position, and magnification for the
weak-deflection regime of the Myers-Perry metric.
Eiroa \cite{Eiroa} and Whisker \cite{Whisker} studied the
bending angle, image position, and magnification for the
strong-deflection regime of the Myers-Perry and tidal
Reissner-Nordstr\"om metrics, respectively.
Still other metrics may prove useful for studying the
approximate lensing properties of braneworld black holes
(cf.\ \cite{GWBD,Whisker}).

We present a thorough study of lensing in braneworld gravity,
including a realistic assessment of prospects for observing
braneworld effects in astrophysical lensing scenarios.  We
focus on weak-deflection lensing for two reasons.  First, the
images that appear in the strong-deflection limit carry
important near-horizon effects but are exceedingly difficult
to observe \cite{virbhadra,petters-SgrA}.  Second, we shall
argue that wave optics will play a crucial role in lensing
tests of braneworld gravity, and wave optics observables are
dominated by the two images that appear in the weak-deflection
regime.  Since the Garriga-Tanaka metric correctly describes
the spacetime geometry in the far-field regime of a braneworld
black hole, we develop the full analytical infrastructure
for weak-deflection lensing in this metric.

We rederive the light bending angle in the Garriga-Tanaka
metric, but express it for the first time in terms of invariant
quantities (Section \ref{sec:bendangle}).
We go beyond the bending angle and use invariant quantities
to compute the observable properties of the lensed images:
positions, magnifications, and time delays
(Sections \ref{sec:posmag}--\ref{sec:tdel}).
We then consider for the first time wave optics effects in
braneworld black hole lensing (Section \ref{sec:waveoptics}).
Finally, we examine a variety of applications of braneworld
lensing.  We show that traditional astrophysical lensing
scenarios will be unable to measure braneworld effects in
the foreseeable future (Section \ref{sec:app1}).  However,
an application of wave optics that we call ``attolensing''
provides exciting opportunities for observing braneworld
effects (Section \ref{sec:app2}).  In particular, we point
out that if primordial braneworld black holes exist and
contribute to the dark matter, they must exist not only
throughout the universe but also within our Solar System.
Attolensing will provide a crucial test of this prediction
of the braneworld model.

\section{Metric and Light Bending Angle}
\label{sec:bendangle}

\begin{figure}
\includegraphics[width=3.2in]{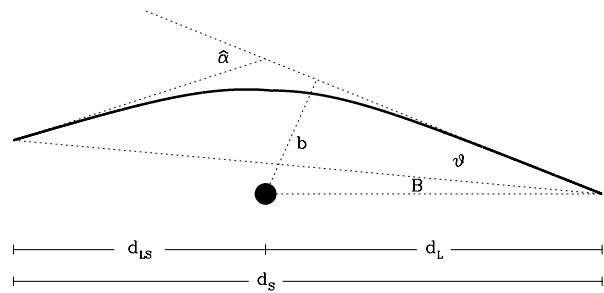}
\caption{
Schematic diagram of the lensing geometry.  Standard quantities
are defined as follows:
$\cb$ is the angular position of the unlensed source;
$\vth$ is the angular position of an image;
$\hat{\alpha}$ is the bending angle;
and $d_L$, $d_S$, and $d_{LS}$ are angular diameter distances
between the observer, lens, and source.  The impact parameter
$b$ is an invariant of the light ray and is related to the
angular image position by $\vth = \sin^{-1}(b/d_L)$.
}
\label{fig:geom}
\end{figure}

We begin by stating our basic assumptions
(cf.\ \cite{paperI,paperII}).  Consider a gravitational lens
with mass $\Mbh$ that is compact, static, and spherically
symmetric, with an asymptotically flat spacetime geometry
sufficiently far from the lens \footnote{Asymptotic flatness
can be generalized to a Robertson-Walker background by using
angular diameter distances and including appropriate redshift
factors; see Section \ref{sec:tdel}.}.  The spacetime is
vacuum outside the lens and flat in the absence of the lens.
We adopt the standard lensing scenario shown in
Figure~\ref{fig:geom}, with the observer and source lying in
the asymptotically flat regime of the spacetime.

The light ray's distance of closest approach $r_0$ and impact
parameter $b$ are both assumed to lie well outside the lens's
gravitational radius $\gravr = G \Mbh/c^2$.  The light bending
angle is assumed to have the following form at lowest order
in $\gravr$:
\beq \label{eq:bangle-ser}
  \hat{\alpha}(b) = A_1 \left(\frac{\gravr}{b}\right)
\left(1 + \frac{B_2}{b^2}\right)
    \ + \ \order{\frac{\gravr}{b}}{2}, 
\eeq
where $A_1$ and $B_2$ are independent of $\gravr$ and $b$.
Since $b$ and $\gravr$ are invariants of the light ray, this
expression for the bending angle is independent of coordinates.  
If $B_2 \ne 0$, the bending angle (\ref{eq:bangle-ser}) cannot
be written as a series in the single parameter $\gravr/b$, which
places this model outside the PPN framework studied in Papers I
and II.  We now show that braneworld black holes do yield
bending angles of this form.

\subsection{Isotropic coordinates}

The Garriga-Tanaka metric is often written as follows in
isotropic coordinates:
\beqa \label{eq:metric1}
  ds^2 &=& - \left( 1 - \frac{2\gravr}{\rr} -
    \frac{4 \gravr \ell^2}{3 \rr^3} \right) dt^2 \\
  &&
    \ + \ \left( 1 + \frac{2\gravr}{\rr} +
    \frac{2 \gravr \ell^2}{3 \rr^3} \right) \left( d\rr^2 + \rr^2\,
    d\Omega^2 \right). \nonumber
\eeqa
This form is valid only in the limit
\beq \label{eq:larger}
  \frac{\gravr}{\rr} \ll 1, \qquad
  \frac{\ell^2}{\rr^2} \ll 1,
\eeq
and the exact metric describing the spacetime geometry around
braneworld black holes is not yet known.  We shall verify
{\em a posteriori} that our lensing solutions satisfy
(\ref{eq:larger}).

The metric (\ref{eq:metric1}) has terms of the form
$\pm 2\gravr/\rr$ as in the weak-deflection limit of general
relativity.  Braneworld effects enter via the $\ell$ terms.
(When $\ell=0$ we recover the standard far-field black hole
metric of general relativity.)  Notice that the $\ell$ terms
factor as $(\gravr/\rr)(\ell^2/\rr^2)$, so we can adopt the
approach of taking Taylor series expansions in $\gravr/\rr$ to
obtain the appropriate weak-deflection limit including the
braneworld terms.  Later we shall consider when it is appropriate
to take series expansions in $\ell/\rr$ as well.

We can think of the metric (\ref{eq:metric1}) more generally
as having the form
\beq \label{eq:metric2}
  ds^2 = -\bar{\cal A}(\rr)\,dt^2 \ + \ \bar{\cal B}(\rr)\,d\rr^2
    \ + \ \bar{\cal C}(\rr)\,\rr^2\,d\Omega^2\,,
\eeq
where the metric functions $\bar{\cal A}(\rr)$ and
$\bar{\cal B}(\rr) = \bar{\cal C}(\rr)$ are readily identified.
For such a metric, the distance of closest approach $\rr_0$ is
related to the impact parameter $b$ by (cf.\ eq.~12 of Paper I)
\beqa
  b &=& \rr_0 \left[\frac{\bar{\cal C}(\rr_0)}{\bar{\cal A}(\rr_0)}
    \right]^{1/2} \nonumber\\
  &=& \rr_0 \left[ 1 + \left(2 + \frac{\ell^2}{\rr_0^2}\right)
    \frac{\gravr}{\rr_0} + \order{\frac{\gravr}{\rr_0}}{2} \right] .
  \label{eq:iso-bofr}
\eeqa
Inverting this relation yields
\beq \label{eq:iso-rofb}
  \rr_0 = b \left[ 1 - \left(2 + \frac{\ell^2}{b^2}\right)
    \frac{\gravr}{b} + \order{\frac{\gravr}{b}}{2} \right] .
\eeq

For a metric of the form (\ref{eq:metric2}), the light bending angle
can be written as
\beq
  \hat{\alpha}(\rr_0) = 2 \int_{\rr_0}^{\infty} \frac{1}{\rr^2} \left[
    \frac{ \bar{\cal A}\ \bar{\cal B} }{ \bar{\cal C}^2/b^2 - 
      \bar{\cal A}\ \bar{\cal C}/\rr^2 }
    \right]^{1/2} \,d\rr \ - \ \pi\,.
\eeq
Plugging in the metric functions, and temporarily replacing $b$ with
$\rr_0$ using (\ref{eq:iso-bofr}), we can write a series expansion
for the integrand,

\begin{widetext}
\beq
  \hat{\alpha}(\rr_0) = 2 \int_{\rr_0}^{\infty} 
   \frac{\rr_0}{\rr\,(\rr^2-\rr_0^2)^{1/2}} \left[
    1 + \frac{2 \rr^2 \rr_0^2 + \ell^2 (\rr^2 + \rr \rr_0 + \rr_0^2)}
      {\rr \rr_0^2 (\rr+\rr_0)}\ \frac{\gravr}{\rr_0}
    + \order{\frac{\gravr}{\rr_0}}{2} \right]\,d\rr 
  \ - \ \pi\,.
\eeq
\end{widetext}

\noindent
Carrying out the integration yields the deflection angle in
terms of the isotropic coordinate distance of closest approach,
\beq
\label{eq:bangle-iso}
  \hat{\alpha}(\rr_0) = 4\,\frac{\gravr}{\rr_0}
    \left( 1 + \frac{\ell^2}{\rr_0^2} \right)
    + \order{\frac{\gravr}{\rr_0}}{2} .
\eeq
This agrees with the bending angle found by Kar \& Sinha \cite{KS}.
However, as emphasized in Paper I, expressions like
(\ref{eq:bangle-iso}) are coordinate dependent and should be
re-expressed in invariant form.  Using (\ref{eq:iso-rofb}) to
rewrite the distance of closest approach $\rr_0$ in terms of the
invariant impact parameter $b$ yields
\beq
\label{eq:bangle-inv}
  \hat{\alpha}(b) = 4\,\frac{\gravr}{b}
    \left( 1 + \frac{\ell^2}{b^2} \right)
    + \order{\frac{\gravr}{b}}{2} .
\eeq
At this order of approximation the form of the bending angle is
the same for $r_0$ and $b$; still, it is important to use the
invariant expression.  Notice that (\ref{eq:bangle-inv}) has the
form assumed in (\ref{eq:bangle-ser}), with $A_1 =4$ and
$B_2 = \ell^2$, which shows that braneworld black hole lensing
lies outside the standard PPN framework. 

\subsection{Standard coordinates}

As a consistency check of (\ref{eq:bangle-inv}), and to connect
with our formalism in Papers I and II, we rederive the bending
angle starting from standard coordinates (or the area gauge
\cite{GWBD}).  In this case we write the metric in the form
\beq \label{eq:std-metric}
  ds^2 = - A(r)\,dt^2 + B(r)\,dr^2 + r^2\,d\Omega^2 .
\eeq
Comparing the $d\Omega^2$ terms in (\ref{eq:metric1}) and
(\ref{eq:std-metric}) shows that the isotropic and standard radial
coordinates are related by
\beq
  r = \rr \left[ 1 + \left(1 + \frac{\ell^2}{3 \rr^2}\right)
    \frac{\gravr}{\rr} + \order{\frac{\gravr}{\rr}}{2} \right] .
\eeq
Inverting this relation yields
\beq
\label{eq:rr-to-b}
  \rr = r \left[ 1 - \left(1 + \frac{\ell^2}{3 r^2}\right)
    \frac{\gravr}{r} + \order{\frac{\gravr}{r}}{2} \right] .
\eeq
Returning to the metric (\ref{eq:metric1}) and changing radial
coordinates yields a metric of the form (\ref{eq:std-metric})
with metric functions
\beqa
  A(r) &=& 1 - 2 \left( 1 + \frac{2\ell^2}{3 r^2} \right) \frac{\gravr}{r}
    + \order{\frac{\gravr}{r}}{2} , \label{eq:std-A} \\
  B(r) &=& 1 + 2 \left( 1 + \frac{ \ell^2}{  r^2} \right) \frac{\gravr}{r}
    + \order{\frac{\gravr}{r}}{2} . \label{eq:std-B}
\eeqa

When the metric is written in the form (\ref{eq:std-metric}), the
standard-coordinate distance of closest approach $r_0$ and invariant
impact parameter $b$ are related by
\beqa
  b &=& \frac{r_0}{\sqrt{A(r_0)}} \nonumber\\
  &=& r_0 \left[ 1 + \left(1 + \frac{2\ell^2}{3 r_0^2}\right)
    \frac{\gravr}{r_0} + \order{\frac{\gravr}{r_0}}{2} \right] , \\
  \label{eq:bofr}
  r_0 &=& b \left[ 1 - \left(1 + \frac{2\ell^2}{3 b^2}\right)
    \frac{\gravr}{b} + \order{\frac{\gravr}{b}}{2} \right] .
  \label{eq:rofb}
\eeqa
The light bending angle can be written as
\beq
  \hat{\alpha}(r_0) = 2 \int_{r_0}^{\infty} \frac{1}{r^2}
    \left[\frac{ \bar{A} \bar{B} }{ 1/b^2 - \bar{A}/r^2 }\right]^{1/2}
    \,dr \ - \ \pi\,.
\eeq
Plugging in the metric functions, and temporarily replacing $b$
with $r_0$ using (\ref{eq:bofr}), we can write a series expansion
for the integrand,

\begin{widetext}
\beqa
  \hat{\alpha}(r_0) &=& 2 \int_{r_0}^{\infty}
   \frac{r_0}{r\,(r^2-r_0^2)^{1/2}} \Biggl[
    1 + \frac{3 r^2 r_0^2 (r^2 + r r_0 + r_0^2)
      + \ell^2 (2 r^4 + 2 r^3 r_0 + 2 r^2 r_0^2 + 3 r r_0^3 + 3 r_0^4)}
      {3 r^3 r_0^2 (r+r_0)}\ \frac{\gravr}{r_0} \nonumber\\
  &&\hspace{1.2in}
    \ + \ \order{\frac{\gravr}{r_0}}{2} \Biggr]\,dr
  \ - \ \pi\,.
\eeqa
\end{widetext}

\noindent
Carrying out the integration yields the deflection angle in terms
of the coordinate distance of closest approach,
\beq
\label{eq:bangle-std}
  \hat{\alpha}(r_0) = 4\,\frac{\gravr}{r_0}
    \left( 1 + \frac{\ell^2}{r_0^2} \right)
    + \order{\frac{\gravr}{r_0}}{2} .
\eeq
Using (\ref{eq:rofb}) to rewrite the distance of closest approach
$r_0$ in terms of the impact parameter $b$ yields
\beq \label{eq:bangle-final}
  \hat{\alpha}(b) = 4\,\frac{\gravr}{b}
    \left( 1 + \frac{\ell^2}{b^2} \right)
    + \order{\frac{\gravr}{b}}{2} .
\eeq
This result agrees with (\ref{eq:bangle-inv}), showing that we
arrive at the same desired invariant bending angle expression
starting from isotropic and standard coordinates.

\section{Image Positions, Magnifications, and Centroid}
\label{sec:posmag}

We now go beyond the bending angle to determine observable
quantities in braneworld black hole lensing.  We examine the
image positions and magnifications in this section, and defer
the time delay to Section \ref{sec:tdel}.  (This analysis
parallels Section IV in Paper I.) We begin with the general
lens equation (cf.\ \reffig{geom}),
\beq \label{eq:lenseq1}
  \tan \cb = \tan \vth - D \ (\tan \vth + \tan (\hat{\alpha} - \vth)) ,
\eeq
where $\cb$ is the angular position of the source,
$\vth = \sin^{-1} (b/d_L)$ is the angular position of the image,
and $D = d_{LS}/d_S$.  We shall see that the lens equation
yields two images in the far-field (weak-deflection) regime,
one on the same side of the lens as the source and the other
on the opposite side.  Following the convention of Papers I
and II, angles describing image positions are taken to be
positive.  This forces the source's angular position to have
different signs: $\cb$ is positive when the image is on the
same side of the lens as the source (as depicted in
\reffig{geom}); while $\cb$ is negative when the image is on
the opposite of the lens from the source.

We now seek an appropriate series expansion of the lens
equation in the weak-deflection limit.  First, we change
variables in light of the fact that lensing quantities
naturally scale with the weak-deflection angular Einstein
ring radius,
\beq \label{eq:vthE}
  \vthE = \sqrt{\frac{4 G \Mbh d_{LS}}{c^2 d_L d_S}}\ .
\eeq
Specifically, we define:
\beq \label{eq:newvar}
  \beta = \frac{\cb}{\vthE}\ , \quad 
  \theta = \frac{\vth}{\vthE}\ , \quad
  \vem = \frac{\vthbh}{\vthE} = \frac{\vthE}{4\,D}\ ,
\eeq
where $\vthbh = \tan^{-1} (\gravr/d_L)$.  In other words,
the quantities $\beta$ and $\theta$ are the scaled angular
positions of the source and image, respectively.  The quantity
$\vem$ represents the angle subtended by the gravitational
radius normalized by the angular Einstein radius, and it
replaces $\gravr/b$ as our expansion parameter.  We also
define an angle associated with the braneworld length scale
$\ell$,
\beq
  \vthl = \tan^{-1}\left(\frac{\ell}{d_L}\right)\ ,
\eeq
and a scaled version of this angle
\beq
  \vel = \frac{\vthl}{\vthE}\ .
\eeq
When it becomes appropriate to consider series expansions
in $\ell/b$ (below), we shall actually use $\vel$ as the
expansion parameter.  The conditions (\ref{eq:larger}) for
validity of the Garriga-Tanaka metric are equivalent to the
conditions $\vem \ll 1$ and $\vel \ll 1$.

With these substitutions, and the bending angle from
(\ref{eq:bangle-final}), the lens equation becomes
\beq \label{eq:lenseq2}
  0 = \left[ - \beta + \theta - \frac{1}{\theta}
    \left( 1 + \frac{\vel^2}{\theta^2} \right) \right] \vem
    + \order{\vem}{2} .
\eeq
To determine braneworld effects on the image positions, we
then postulate that the position can be expanded in the form
\beq
  \theta = \theta_0 \ + \ \theta_1\,\vel \ + \
    \theta_2\,\vel^2 \ + \ \order{\vel}{3} .
\eeq
Plugging this into (\ref{eq:lenseq2}) yields
\beqa \label{eq:lenseq3}
  0 &=& \left( -\beta + \theta_0 - \frac{1}{\theta_0} \right)
    \ + \ \left( 1 + \frac{1}{\theta_0^2} \right) \theta_1\,\vel \\
  &&
    \ + \ \frac{(1+\theta_0^2) \theta_0 \theta_2 - 1 - \theta_1^2}{\theta_0^3}\ 
          \vel^2
    \ + \ \order{\vel}{3} . \nonumber
\eeqa
This is the desired series expansion of the lens equation.

The zeroth-order term in (\ref{eq:lenseq3}) is just the usual
lens equation for the weak-deflection limit of general relativity,
whose solution is
\beq
\label{eq:theta0} 
  \theta_0 = \frac{1}{2} \left( \sqrt{\beta^2+4} + \beta \right) .
\eeq
We neglect the negative solution because by convention angles
describing image positions are positive.  The positive-parity
image $\theta_0^+$, which lies on the same side of the lens as
the source, is found by using $\beta > 0$.  The negative-parity
image $\theta_0^-$ is then found by using $\beta < 0$.
Explicitly,
\beq \label{eq:theta0pm} 
  \theta_0^\pm = \frac{1}{2} \left(\sqrt{4 + \beta^2} \pm |\beta|\right) .
\eeq

The first-order term in (\ref{eq:lenseq3}) can be satisfied
only if $\theta_1 = 0$.  In other words, {\em there is no
braneworld correction to the lensed image positions at first
order in $\vel \sim \ell/b$}.  This is not surprising, since
braneworld effects enter the metric at order $(\ell/r)^2$.

The second-order term in (\ref{eq:lenseq3}) is satisfied if
\beq
  \theta_2 = \frac{1}{\theta_0 (1+\theta_0^2)}\ .
\eeq
Thus, the full expression for the image position in braneworld
gravity is
\beq \label{eq:pos}
  \theta = \theta_0 \ + \ \frac{\vel^2}{\theta_0 (1+\theta_0^2)}
    \ + \ \order{\vel}{3}.
\eeq
Rewriting $\theta_0$ in terms of $\beta$ using (\ref{eq:theta0pm}),
we can express the image positions in terms of the source position
as
\beqa \label{eq:pospm}
  \theta^\pm &=& \frac{1}{2} \left(\sqrt{4+\beta^2} \pm |\beta|\right) \\
  &&
    \ + \ \frac{1}{2} \left( \frac{2+\beta^2}{\sqrt{4+\beta^2}}
      \mp |\beta| \right)\,\vel^2
    \ + \ \order{\vel}{3}. \nonumber
\eeqa
The coefficient of $\vel^2$ is positive for all value of $\beta$,
which means that braneworld effects push both the positive- and
negative-parity images {\em farther} away from the lens (relative
to the results for general relativity).

By spherical symmetry, the signed magnification $\mu$ of a
lensed image at angular position $\vth$ is
\beq
  \mu(\vth) = \left[\frac{\sin \cb (\vth)}{\sin \vth} \
    \frac{d \cb(\vth)}{d \vth} \right]^{-1} .
\eeq
After changing to our scaled variables, we first make a Taylor
series expansion in $\vem$:
\beq
  \mu = \frac{\theta^8}{(\theta^4+\theta^2+3\vel^2)
    (\theta^4-\theta^2-\vel^2)} \ + \ \order{\vem}{} .
\eeq
Now using (\ref{eq:pos}) for the image position, we find
\beq \label{eq:mag}
  \mu  =  \frac{\theta_0^4}{\theta_0^4-1}
    \ - \ \frac{2\, \theta_0^4}{(\theta_0^4-1)(\theta_0^2+1)^3}\,\vel^2
    \ + \ \order{\vel}{3} .
\eeq 
The absolute magnifications in terms of the source position
are given by
\beqa \label{eq:magpm}
  |\mu^\pm| &=& \frac{1}{2} \left( \frac{2+\beta^2}{|\beta| \sqrt{4+\beta^2}} 
    \pm 1 \right) \\
  &&
    \ - \ \frac{2}{|\beta|(4+\beta^2)^{3/2}}\,\vel^2
    \ + \ \order{\vel}{3} , \nonumber
\eeqa
where $\mu^- < 0$.   Observe that at lowest order the braneworld
magnifications have $\mu^+ + \mu^- = 1$.  This is identical to
the lowest-order universal magnification relation in eq.~(36)
of Paper II.  It is interesting to see that braneworld gravity
obeys the relation originally derived for PPN models, even though
it lies outside the PPN framework.

In cases where the two images cannot be separately resolved
(such as microlensing \cite{Saas-Fee,micro-pac,GP}), the useful
quantities are the total magnification and the
magnification-weighted centroid position.  The total
magnification can be written as
\beqa \label{eq:mutot}
  \mu_{\rm tot} &\equiv& |\mu^+| + |\mu^-| \\
  &=& \frac{2+\beta^2}{|\beta|\sqrt{4+\beta^2}} \ - \
    \frac{4}{|\beta|(4+\beta^2)^{3/2}}\,\vel^2
    \ + \ \order{\vel}{3} . \nonumber
\eeqa
The magnification-weighted centroid position can be written as
\beqa \label{eq:centroid}
  \Theta_{\rm cent} &\equiv& \frac{\theta^+ |\mu^+| - \theta^- |\mu^-|}
    {|\mu^+| + |\mu^-|} \\
  &=& \frac{|\beta| \,(3+\beta^2)}{2+\beta^2} \ + \
    \frac{2|\beta|}{(2+\beta^2)^2}\,\vel^2
    \ + \ \order{\vel}{3} . \nonumber
\eeqa
We see that braneworld effects decrease the total magnification
and push the centroid farther from the lens (compared with the
results for general relativity).

\section{Time Delay}
\label{sec:tdel}

We now derive the lensing time delay (in parallel with Section V
of Paper I).  We first focus on a spacetime that is static and
asymptotically flat, and discuss the generalization to a curved
universe cosmology at the end of this section.

Let $R_{\rm src}$ and $R_{\rm obs}$ be the radial coordinates of
the source and observer, respectively.  From geometry relative
to the flat metric of the distant observer (who is assumed to be
at rest in the natural coordinates of the metric
eq.~\ref{eq:std-metric}), we can work out (see \reffig{geom})
\beq \label{eq:Rdef}
  R_{\rm obs} = d_L\,, \qquad
  R_{\rm src} = \left( d_{LS}^2 + d_S^2 \tan^2 \cb \right)^{1/2} .
\eeq
The radial distances are very nearly the same as angular diameter
distances since the source and observer are in the asymptotically
flat region of the spacetime.  In other words, the distortions
in distances near the black hole are assumed to have little
impact on the total flat metric distance from the compact body
to the observer or source.

In the absence of the lens the light ray would travel along a
linear path from the source to the observer with length
$d_S/\cos\cb$.  The time delay $\tau$ is the difference between
the light travel time for the actual ray, and the travel time for
the straight line the ray would have taken in the absence of the
lens.  This can be written as
\beq
  c \tau = {\cal T}(R_{\rm src}) + {\cal T}(R_{\rm obs})
    - \frac{d_S}{\cos\cb}\ ,
\eeq
with (see eq.~91 of Paper I)
\beq
  {\cal T}(R) = \frac{1}{b} \, \int_{r_0}^{R} \frac{1}{A(r)} \
    \sqrt{\frac{A(r) \, B(r) }{1/b^2 \ - \ A(r)/r^2}}\ dr\,.
\eeq
Using the metric functions (\ref{eq:std-A}) and (\ref{eq:std-B}),
we find for braneworld gravity

\begin{widetext}
\beqa
  {\cal T}(R) &=& \int_{r_0}^{R} \frac{r}{\sqrt{r^2-r_0^2}} \Biggl[
    1 + \frac{3 r^2 r_0 (2 r + 3 r_0) + \ell^2 (2 r^2 + 7 r r_0 + 7 r_0^2)}
      {3 r^3 r_0 (r+r_0)}\ \frac{\gravr}{r_0} 
    + \order{\frac{\gravr}{r_0}}{2} \Biggr]\,dr , \nonumber\\
  &=& \sqrt{R^2-r_0^2} + \left[ 
    2 \ln \left(\frac{R+\sqrt{R^2-r_0^2}}{r_0}\right)
    + \sqrt{R-r_0} \left( 1 + \ell^2\,\frac{9 R + 7 r_0}{3 R r_0^2} \right)
    \right] 
  \ + \order{\frac{\gravr}{r_0}}{2} .
\eeqa
\end{widetext}

\noindent
To obtain a coordinate-invariant expression, we could replace $r_0$
with $b$ using (\ref{eq:rofb}).  We would then want to take a series
expansion in $b/R$ as well as $\gravr/b$, because they are of the
same order (see Section V of Paper I).

It is simpler to proceed directly to the time delay expressed in
terms of our scaled angular variables.  We compute
${\cal T}(R_{\rm src})$ and ${\cal T}(R_{\rm obs})$ using the
radii from (\ref{eq:Rdef}).  We change to angular variables using
$b = d_L \sin\vth$, and then reintroduce the scaled angular
variables $\theta$ and $\beta$ defined in (\ref{eq:newvar}).  We
work to lowest order in $\vem$, and then take a Taylor series in
the dimensionless braneworld parameter $\vel$.  The result is
\beq
  \frac{\tau}{\tau_E} =
    \frac{1}{2} \left[ 1 + \beta^2 - \theta_0^2 - \ln \left(
    \frac{d_L\,\theta_0^2\,\vthE^2}{ 4\,d_{LS}} \right) \right]
  \ + \ \frac{\vel^2}{2\theta_0^2}
  \ + \ \order{\vel}{3} ,
\eeq
where the natural time scale is
\beq \label{eq:tauE}
  \tau_E \equiv \frac{d_L d_S}{c\,d_{LS}}\ \vthE^2
  = 4\,\frac{\gravr}{c}\ .
\eeq
Note that we have used (\ref{eq:pos}) for the image position to
obtain an expression written in terms of $\theta_0$, the image
position in the weak-deflection limit of general relativity.

More interesting than the individual time delays is the differential
delay $\Delta\tau = \tau^- - \tau^+$ between the negative- and 
positive-parity images:
\beqa
  \frac{\Delta\tau}{\tau_E} &=& \left[ \frac{ (\theta_0^+)^2 -
    (\theta_0^-)^2 }{2} + \ln\frac{\theta_0^+}{\theta_0^-} \right] \nonumber\\
  &&
    \ + \ \frac{\vel^2}{2} \Bigl[ (\theta_0^-)^{-2} - (\theta_0^+)^{-2} \Bigr]
    \ + \ \order{\vel}{3} , \\
  &=& \left[ \frac{1}{2}\,|\beta| \sqrt{4+\beta^2} +
    \ln\left(\frac{\sqrt{4+\beta^2}+|\beta|}{\sqrt{4+\beta^2}-|\beta|}\right)
    \right] \nonumber\\
  &&
    \ + \ \frac{\vel^2}{2}\,|\beta| \sqrt{4+\beta^2}
    \ + \ \order{\vel}{3} .
\eeqa
The first expression is written in terms of the image positions
$\theta_0^\pm$, while the second is written in terms of the
source position.  In each case, the order unity term recovers
the familiar time delay in the weak-deflection limit of general
relativity.  To simplify the notation below, we write the
differential time delay as
\beq \label{eq:tdel}
  \frac{\Delta\tau}{\tau_E} 
  = \Lambda_0 + \Lambda_1 + \Lambda_0 \, \vel^2 + \order{\vel}{3} ,
\eeq
where 
\beq \label{eq:Lambda01}
  \Lambda_0 = \frac{1}{2}\,|\beta| \sqrt{4+\beta^2}\ , \quad
  \Lambda_1 = \ln\left(\frac{\sqrt{4+\beta^2}+|\beta|}
    {\sqrt{4+\beta^2}-|\beta|}\right) .
\eeq

It is straightforward to place this analysis in a background
universe that is curved and expanding.  We are working in the
limit that the impact parameter is small compared to the
distances between the observer, lens, and source.  Thus, the
light path is determined by the background cosmology for all
but the tiny fraction of the path when it is near the lens.  
(See \cite{SEF,PLW} for further discussion.)  All of the previous
analysis holds if we interpret $d_L$, $d_S$, and $d_{LS}$ as
cosmological angular diameter distances, and we modify the
natural time scale from (\ref{eq:tauE}) to
\beq \label{eq:tauE2}
  \tau_E \equiv (1+z_L)\ \frac{d_L d_S}{c\,d_{LS}}\ \vthE^2
  = 4\,(1+z_L)\,\frac{\gravr}{c}\ ,
\eeq
where $z_L$ is the cosmological redshift of the lens.

\section{Semi-Classical Interference Optics}
\label{sec:waveoptics}

The foregoing analysis applies in the limit of geometric optics.
We now examine how the wave nature of light can affect gravitational
lensing observables.  Wave optics have been studied for standard
weak-deflection lensing in general relativity (see the review by
Nakamura \& Deguchi \cite{nakamura}, and Sections 4.7 and Chapter 7
of Schneider et al.\ \cite{SEF}), but to our knowledge have not
been treated before in the braneworld lensing literature.

In analogy with the Young double-slit experiment, we may consider
that light waves from the positive- and negative-parity images
interact to produce an interference pattern at the observer.  To
quantify different regimes, suppose a pointlike light source emits
a monochromatic, spherical light wave of period $T = 2\pi/\omega$
or wavelength $\lambda = c\,T$, which is lensed by a braneworld
black hole.  Geometric optics apply in the limit $T \to 0$.  If
the period is finite but small compared with the time delay between
the images ($T \ll \Delta\tau$), then we are in the
{\em semi-classical limit} (cf.\ \cite{nakamura,ulmer}).  This is
the regime we shall investigate.  Larger periods
$T \gtrsim \Delta\tau$ lead to the physical optics limit.  In this
regime, all regions of the lens plane contribute to interference
effects.  The analysis therefore requires knowledge of the time
delay function across the entire lens plane, which in turn
requires knowledge of the metric at all such positions.  However,
the full metric for braneworld black holes is not yet known, which
means that we cannot give a complete wave optics treatment of lensing
in braneworld gravity at this time.

There is a well-established connection between the geometric optics
and semi-classical limits.  If geometric optics predict a set of
images with magnifications $\mu_i$ and time delays $\Delta\tau_{ij}$,
then the total magnification in the semi-classical limit is given
as follows (e.g., Sections 4.7 and 7.1 of \cite{SEF}, and Section 2
of \cite{nakamura}):
\beq \label{eq:mm-general}
  \mm = \sum_j |\mu_j|
    \ + \ 2 \sum_{i<j} |\mu_i \mu_j |^{1/2}\,
      \cos\left[ \omega \Delta\tau_{ji} - (n_j-n_i) \frac{\pi}{2} \right] .
\eeq
where $n_k = 0,1,2$ depending on whether the $k$-th image is a
minimum, saddle, or maximum, respectively.  The first sum gives
the total magnification in the geometric optics limit, while
the second captures the corrections due to semi-classical
interference between the images.  There is a phase factor of
$\omega \Delta\tau_{ij}$ from the lensing time delay, and an
additional factor of $(n_j-n_i) \pi/2$ from differences in the
phases of the images upon exiting the lens plane.

Lensing by a black hole produces a positive  and negative
image pair, so we have $n_{12} = 1 - 0 =1$ and
$\Delta\tau_{12} =  \tau^- - \tau^+$.  The total magnification
in the semi-classical limit can then be written as 
\beq \label{eq:mm}
  \mm = |\mu^+| \ + \ |\mu^-| \ + \
    2\,|\mu^+ \mu^-|^{1/2}\ \sin(\omega \Delta\tau)\,.
\eeq
The sine term creates a series of bright and dark fringes in the
magnification as a function of energy or wavelength.  Bright
fringes (maxima in $\mm$) occur when the phase difference is
$\omega \Delta\tau = 2j\pi + \pi/2$ for integer $j$, which
correspond to photon energies
\beq
  E^{\rm br}_j = (j + 1/4)\ \frac{h}{\Delta \tau}\ ,
  \qquad j = 0,1,2,\dots
\eeq
Dark fringes (minima in $\mm$) occur when the phase difference
is $\omega \Delta\tau = (2j+1)\pi + \pi/2$, or energies
\beq
  E^{\rm dk}_j = (j + 3/4) \ \frac{h}{\Delta \tau}\ ,
  \qquad j = 0,1,2,\dots.
\eeq
Note that the energy spacing between adjacent bright fringes,
or adjacent dark fringes, is always $\Delta E = h/\Delta\tau$,
independent of $j$.

For a braneworld black hole, the magnification sum
$|\mu^+| + |\mu^-|$ is given in (\ref{eq:mutot}), and the
magnification product is
\beq
  |\mu^+ \mu^-|^{1/2} = \frac{1}{|\beta| \sqrt{4+\beta^2}}
    \ - \ \frac{2+\beta^2}{|\beta|(4+\beta^2)^{3/2}}\,\vel^2
    \ + \ \order{\vel}{3} .
\eeq
Consequently, the semi-classical total magnification is given
as follows in terms of the source position:
\beqa \label{eq:mm2}
  \mm &=& \frac{\sqrt{4 + \beta^2}}{|\beta|} \Biggl[
    \frac{2 + \beta^2 + 2 \sin(\omega \Delta \tau)}{4 + \beta^2} \\
  &&
    \ - \ \frac{4 + 2 (2 + \beta^2) \sin(\omega \Delta \tau)}
      {(4 + \beta^2)^2}\ \vel^2 \ + \ \order{\vel}{3}
  \Biggr] . \nonumber
\eeqa
Using the time delay from (\ref{eq:tdel}), we find that the
bright fringes have magnification
\beq \label{eq:mbri}
  \mm^{\rm br} = \frac{\sqrt{4+\beta^2}}{|\beta|} \left[
    1 \ - \ \frac{2}{4+\beta^2}\,\vel^2 \ + \ \order{\vel}{3} \right] ,
\eeq
and are located at energies
\beq \label{eq:Ebri}
  E_j^{\rm br} = \frac{(j + 1/4) h}{(\Lambda_0+\Lambda_1) \tau_E}
    \left[ 1 \ - \ \frac{\Lambda_0}{\Lambda_0+\Lambda_1}\ \vel^2
    \ + \ \order{\vel}{3} \right] ,
\eeq
while the dark fringes have magnification
\beq \label{eq:mdrk}
  \mm^{\rm dk} = \frac{|\beta|}{\sqrt{4+\beta^2}} \left[
    1 \ + \ \frac{2}{4+\beta^2}\,\vel^2 \ + \ \order{\vel}{3} \right] ,
\eeq
and are located at energies
\beq \label{eq:Edrk}
  E_j^{\rm dk} = \frac{(j + 3/4) h}{(\Lambda_0+\Lambda_1) \tau_E}
    \left[ 1 \ - \ \frac{\Lambda_0}{\Lambda_0+\Lambda_1}\ \vel^2
    \ + \ \order{\vel}{3} \right] ,
\eeq
The absolute peak-to-trough distance between bright and dark fringes
is
\beq
 \mm^{\rm br} - \mm^{\rm dk} = \frac{4}{|\beta|\sqrt{4+\beta^2}} \left[
    1 \ - \ \frac{2+\beta^2}{4+\beta^2}\,\vel^2 \ + \ \order{\vel}{3} \right] ,
\eeq
while the fractional difference is
\beq \label{eq:Mvis}
  \frac{\mm^{\rm br} - \mm^{\rm dk}}{\mm^{\rm br} + \mm^{\rm dk}}
  = \frac{2}{2+\beta^2} \left[ 1 \ - \ \frac{\beta^2}{2+\beta^2}\,\vel^2
    \ + \ \order{\vel}{3} \right] .
\eeq
We see that braneworld effects (from the extra dimension of space)
shift the interference fringes to lower energies, and contract
the energy spacing $E^{\rm dk}_j -  E^{\rm br}_j$.
Braneworld effects also reduce the magnification of the bright
fringes and increase the magnification of the dark fringes,
which reduces the peak-to-trough distance in the fringe pattern.

\section{Large Braneworld Black Holes in Traditional Lensing Scenarios}
\label{sec:app1}

We have seen that all the braneworld corrections to 
weak-deflection black hole lensing scale with
\beq \label{eq:small-M}
  \vel^2 = \left[\frac{\tan^{-1}(\ell/d_L)}{\vthE}\right]^2
  \approx \frac{\ell^2 c^2}{4 G \Mbh}\ \frac{d_S}{d_L\,d_{LS}}
  = \frac{\ell^2}{4\gravr}\ \frac{d_S}{d_L\,d_{LS}}\ ,
\eeq
where we used $\tan^{-1}(\ell/d_L) \approx \ell/d_L$ and the
definition of $\vthE$ from (\ref{eq:vthE}).  Notice that
$\vel$ is given by the ratio of the braneworld scale $\ell$ to
the geometric mean of the gravitational scale $\gravr$ and the
astrophysical distance scale.

There are two ways to maximize braneworld effects.  One is to
consider a small black hole mass, since $\vel^2 \propto \gravr^{-1}$.
The other possibility is to make either $d_L$ or $d_{LS}$ small.
Notice that if $d_L \ll d_S$ then $d_{LS} \approx d_S$ and the
combination of distances reduces to $1/d_L$.  Alternatively, if
$d_{LS} \ll d_S$ then $d_L \approx d_S$ and the combination of
distances reduces to $1/d_{LS}$.  We can combine these two
possibilities and write
\beq \label{eq:small-d}
  \vel^2 \approx \frac{\ell^2}{4\,\gravr\,\dmin}\ ,
  \qquad \dmin \equiv \min(d_L,d_{LS}) \ll d_S\,.
\eeq
Remarkably, $\vel^2$ does not depend on the distance to the
source if $d_S \gg \dmin$.    

We now examine traditional astrophysical scenarios in which
lensing is or may be observed, and assess whether braneworld
effects are likely to be detectable.  We shall see that all of
the scenarios have $\vel \ll 1$, so the condition (\ref{eq:larger})
is satisfied and our use of the Garriga-Tanaka metric is valid.

Primoridal braneworld black holes may grow to supermassive
scales by accreting dark energy \cite{bean}.  Therefore let us
consider lensing by the supermassive black hole at the center of
our Galaxy.  The black hole has a mass of
$\Mbh = (3.6\pm0.2)\E{6}\,M_\odot$ \cite{ghez}, corresponding
to $\gravr = 5.3\E{11}\mbox{ cm} = 1.7\E{-7}\mbox{ pc}$, and a
distance $d_L = 7.9 \pm 0.4$ kpc \cite{eisen}.  Assuming that
the lens-source distance $d_{LS}$ is much smaller than the
observer-lens distance $d_L$, eq.~(\ref{eq:small-d}) yields
\beq
  \vel^2 \approx 6.1\E{-35} \times
    \left(\frac{\ell}{0.2\mbox{ mm}}\right)^2
    \left(\frac{d_{LS}}{\mbox{pc}}\right)^{-1} .
\eeq
It would be very challenging to measure braneworld effects in
lensing by the Galactic black hole.

To increase $\vel^2$ we need to lower the black hole mass.
Considering stellar-mass black holes brings us into the realm of
microlensing (e.g., \cite{Saas-Fee,micro-pac,GP}).  In Galactic
microlensing the typical distances between observer, lens, and
source are kiloparsecs, so eq.~(\ref{eq:small-M}) yields
\beq
  \vel^2 \approx 2.2\E{-31} \times
    \left(\frac{\ell}{0.2\mbox{ mm}}\right)^{2}
    \left(\frac{\Mbh}{M_\odot}\right)^{-1}
    \left(\frac{d_S}{d_L\,d_{LS}}\right) ,
\eeq
where the distances $d_i$ are in kpc.  Compared with the example
of the Galactic supermassive black hole, we have gained six
orders of magnitude in mass but lost three in distance, so the
braneworld effects are still very small.

An intriguing microlensing system is the binary pulsar
J0737$-$3039 \cite{J0737}.  The binary orbit is seen nearly
edge-on, so that when one neutron star passes behind the other
in projection there may be significant lensing effects
\cite{rafikov}.  Although the lens in this case is not a black
hole, it is still a compact object to which our formalism
applies.  The Einstein radius in this system is 2550 km while
the gravitational radius is $\gravr = 1.85$ km \cite{rafikov},
so the weak-deflection regime applies.  The lens-source distance
is given by the semimajor axis of the orbit,
$d_{LS} \approx a = 8.79\E{5}$ km, and is much smaller than the
distance to the lens, so eq.~(\ref{eq:small-d}) yields
\beq
  \vel^2 \approx 6.1\E{-21} \times
    \left(\frac{\ell}{0.2\mbox{ mm}}\right)^{2} .
\eeq
Braneworld effects are still negligible.

We conclude that in traditional lensing scenarios with black
holes that are stellar mass or larger, the lensing scale is
simply much too large to provide an effective probe of
braneworld effects on the scale $\ell \lesssim 0.2$ mm.

\section{Attolensing by Primordial Braneworld Black Holes}
\label{sec:app2}

To obtain larger braneworld effects, we need to consider even
smaller black hole masses.  In the braneworld model, black holes
may be created in the early universe and survive to the present
day with masses as small as $\sim 1 \mbox{ kg} \sim 10^{-30}\,M_\odot$
\cite{primordial-3,MajMuk1}.  Primordial braneworld black holes
could be detectable via lensing in the wave optics limit, in a
phenomenon we call attolensing.

Attolensing is descended from the concept of femtolensing in
general relativity, introduced by Gould \cite{gould}.  A black
hole of mass $\Mbh \sim 10^{-16}\,M_\odot$ placed at a
cosmological distance creates images with an angular spacing of
order a femto-arcsecond.  Femtolensing could produce observable
interference fringes in the energy spectrum of a gamma-ray
burst at energies in the range of keV to MeV.  Various aspects
of the interference patterns produced by femtolensing have been
studied for general relativity
\cite{stanek,jarosynski,ulmer,gouldgaudi}.

Likewise, attolensing by primordial braneworld black holes
would be observable through wave optics effects.  There are,
however, some notable differences between femtolensing in
general relativity and attolensing in braneworld gravity.
First, at a given black hole mass, braneworld gravity and
general relativity predict slightly different interference
patterns (see Section \ref{sec:waveoptics}).  Second,
braneworld gravity allows lower mass black holes to survive
to the present day (compared with GR); as a result, there are
energy scales at which interference effects would be observable
for braneworld gravity but not general relativity.  Third, as
a point of terminology, Gould's term referred to the scale of
the angular image separation.  By contrast, we choose a term
that denotes the mass scale, which for wave optics is much
more fundamental than the angular image separation.  As noted
above, primordial braneworld black holes can have an enormous
range of masses, down to $\sim\!10^{-30}\,M_\odot$.  In
quantitative examples we take $M \sim 10^{-18}\,M_\odot$ to
be illustrative (hence ``attolensing''), but always quote the
mass scaling.

\begin{figure}
\includegraphics[width=3.0in]{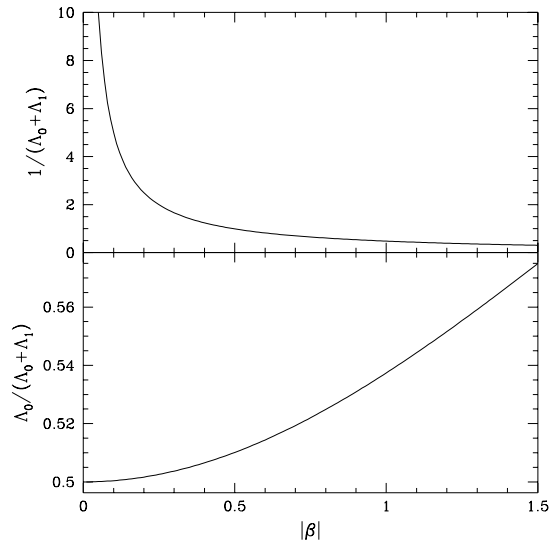}
\caption{
Dimensionless factors in the energies of attolensing interference
fringes (see eq.~\ref{eq:Ebri2}), which depend on the scaled source
position $\beta$.
}
\label{fig:Lambda}
\end{figure}

The gravitational radius and time scale for attolensing are
\beqa
  \gravr &=& 1.5\E{-13} \times
    \left(\frac{\Mbh}{10^{-18}\,M_\odot}\right) \mbox{ cm} , \\
  \tau_E &=& 2.0\E{-23} \times (1+z_L)
    \left(\frac{\Mbh}{10^{-18}\,M_\odot}\right) \mbox{ s} ,
\eeqa
where to be general we consider that the lens may have
cosmological redshift $z_L$.  The bright interference fringes
appear at energies (see eq.~\ref{eq:Ebri})
\beqa \label{eq:Ebri2}
  E_j^{\rm br} &=& 210 \times \frac{j+1/4}{1+z_L}\ \frac{1}{\Lambda_0+\Lambda_1}
    \left(\frac{\Mbh}{10^{-18}\,M_\odot}\right)^{-1} \times \nonumber\\
  &&
    \left[ 1 \ - \ \frac{\Lambda_0}{\Lambda_0+\Lambda_1}\,\vel^2
      \ + \ \order{\vel}{3} \right] \mbox{ MeV} .
\eeqa
The dark fringes are found by replacing $j+1/4$ with $j+3/4$
(cf.\ eq.~\ref{eq:Edrk}).  The dimensionless factors involving
$\Lambda_0$ and $\Lambda_1$ depend on the source position as
shown in \reffig{Lambda}.  Braneworld effects introduce a
fractional shift in the fringe energies (relative to GR) that
is given by $\vel^2$ times a factor that is approximately 0.5
for all relevant source positions.

\subsection{Braneworld Black Holes in the Solar System}

Since the amplitude of braneworld effects decreases with the
distance to the lens (see eq.~\ref{eq:small-d}), we can
estimate the maximum realistic effects by considering primordial
braneworld black holes that are as close as possible to Earth.
To estimate how close this might be, let $\rho_\bh$ be the mean
mass density in primordial braneworld black holes in the Solar
neighborhood.  We might take this to be some fraction $f_\bh$ of
the density $\rho_\dm$ of dark matter in our region of the Galaxy.
If all braneworld black holes have the same mass $\Mbh$, then
their number density is $n_\bh = \rho_\bh/\Mbh$, so the typical
distance between black holes is
\beq
  d \sim n_\bh^{-1/3} = \left(\frac{f_\bh \rho_\dm}{\Mbh}\right)^{-1/3} .
\eeq
Detailed modeling of our Galaxy indicate that the density of
dark matter in the Solar neighborhood is
$\rho_\dm = (0.011\pm0.005)\,M_\odot\mbox{ pc}^{-3}$ \cite{olling}.
This dark matter density yields
\beq
  d_L \sim 0.93 \times f_\bh^{-1/3}
    \left(\frac{\Mbh}{10^{-18}\,M_\odot}\right)^{1/3} \mbox{ AU} .
\eeq
Note that the natural unit here is Astronomical Units.  In
other words, if $\Mbh/M_\odot \lesssim 10^{-13}\,f_\bh$ then
the nearest primordial black holes reside within our Solar
System!

Put another way, a simple estimate of the total mass in
primordial black holes within the Solar System is
\beq \label{eq:SS-Mtot}
  M_\bh \sim \frac{4}{3} \pi R_P^3\,\rho_\bh
  \sim 3.3\E{-13}\ f_\bh\ M_\odot\,,
\eeq
where $R_P \approx 40$ AU is the radius of Pluto's orbit.
This corresponds to a total number
\beq
  N_\bh \sim 3.3\E{5} \times f_\bh
    \left(\frac{M}{10^{-18}\,M_\odot}\right)^{-1}
\eeq
of primordial braneworld black holes in the volume interior
to Pluto's orbit.

The angular Einstein radius and braneworld correction scale
for a primordial braneworld black hole in the Solar System
are
\beqa
  \vthE &\sim& 4.1\E{-8} \times \\
  &&
    \left(\frac{\Mbh}{10^{-18}\,M_\odot}\right)^{1/2}
    \left(\frac{d_L}{\mbox{AU}}\right)^{-1/2} \mbox{arcsec}, \nonumber\\
  \vel^2 &\sim& 4.5\E{-5} \times \label{eq:atto-2}\\
  &&
    \left(\frac{\ell}{0.2\mbox{ mm}}\right)^{2}
    \left(\frac{\Mbh}{10^{-18}\,M_\odot}\right)^{-1}
    \left(\frac{d_L}{\mbox{AU}}\right)^{-1} . \nonumber
\eeqa
In other words, braneworld corrections for lensing by
primordial braneworld black holes in the Solar System are
small but not absurdly so.

The primordial braneworld black holes we are considering are
much less massive than asteroids.  In fact, the total mass in
(\ref{eq:SS-Mtot}) is smaller than many large asteroids
\cite{hilton}, and is well within upper limits on dark matter
in the Solar System derived from Solar System dynamics
\cite{anderson,gron}.  Our estimate of the mass and number of
primordial black holes in the Solar System is based on the
assumption that the dark matter is distributed uniformly in
the Solar neighborhood, which seems plausible because stars
and the star formation process are inefficient at capturing
dark matter \cite{anderson}.

\subsection{Braneworld Black Holes in the Cosmos}

Rather than considering braneworld black holes in our Solar
System, we may imagine them spread throughout the universe.
The angular Einstein radius and braneworld correction scale
for a primordial braneworld black hole at a cosmological distance
are
\beqa
  \vthE &\sim& 2.9\E{-15} \times \\
  &&
    \left(\frac{\Mbh}{10^{-18}\,M_\odot}\right)^{1/2}
    \left(\frac{d_{LS}}{d_L\,d_S}\right)^{1/2} \mbox{arcsec}, \nonumber\\
  \vel^2 &\sim& 2.2\E{-19} \times \\
  &&
    \left(\frac{\ell}{0.2\mbox{ mm}}\right)^{2}
    \left(\frac{\Mbh}{10^{-18}\,M_\odot}\right)^{-1}
    \left(\frac{d_S}{d_L\,d_{LS}}\right) , \nonumber
\eeqa
where the distances are all in Gpc.  In other words, both the
angular image separation and the braneworld corrections would
be difficult to measure for cosmological primordial braneworld
black holes.  However, the interference fringes are still very
similar to the Solar System case; they shift only by the order
unity factor $1+z_L$ (see eq.~\ref{eq:Ebri2}).

An obvious question is whether the probability of cosmological
attolensing is high enough to be interesting.  Suppose that
primordial braneworld black holes contribute a fraction $\Obh$
of the total energy density of the universe.  The lensing
optical depth --- defined to be the fraction of the sky covered
by Einstein radii, which is useful not only for traditional
lensing but for attolensing as well --- would then be $\tau$
such that $\tau/\Obh \gtrsim 0.1$, with the precise coefficient
determined by the distribution of source redshifts
\cite{gould,PressGunn}.  The optical depth is independent of
the distribution of primordial black hole masses, and is nearly
equal to the attolensing probability \cite{holz}.  In other
words, if primordial black holes contribute a cosmologically
significant fraction of matter, then the attolensing probability
is not very small.

\subsection{On Observing Attolensing}

For our example black hole of mass $\Mbh = 10^{-18}\,M_\odot$,
eq.~(\ref{eq:Ebri2}) indicates that semi-classical wave optics
effects would be seen at energies of tens to hundreds of MeV.
This energy range will soon be accessible with the GLAST
satellite, scheduled for launch in 2007 \cite{GLAST}.  (At
present, energies up to 8 MeV can be observed with the INTEGRAL
satellite \cite{INTEGRAL}.  There may be important attolensing
effects that could be seen at these lower energies, but they
would involve physical optics effects and it is not yet possible
to examine braneworld black hole lensing in the physical optics
regime [see Section \ref{sec:waveoptics}].)

\begin{figure}
\includegraphics[width=3.0in]{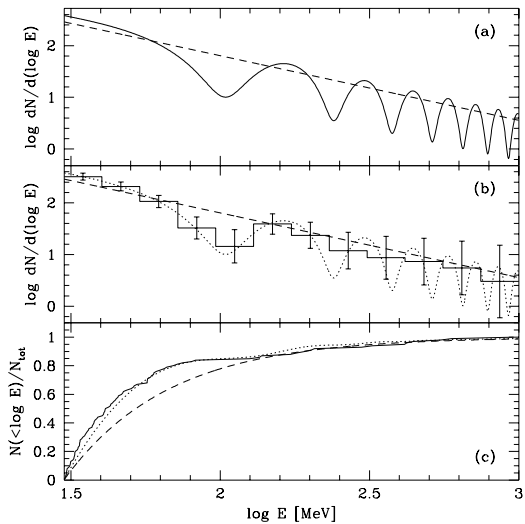}
\caption{
Sample energy spectrum of a gamma-ray burst attolensed by a
braneworld black hole with mass $\Mbh = 10^{-18}\,M_\odot$,
for a configuration in which the scaled source position is
$\beta = 0.75$.
(a) The dashed lines shows the intrinsic energy spectrum,
while the dotted line shows the attolensed spectrum (both
normalized to 100 photons).  We follow the convention for
high-energy spectra and plot the logarithm of the number
of photons per logarithmic energy interval.
(b) The dashed and dotted lines reproduce the ideal spectra
from panel (a).  The histogram shows a sample ``observation''
as it might be seen by GLAST.  The errorbars indicate the
uncertainty in the ``observed'' spectrum if the telescope
records 100 photons with energies between 30 MeV and 1 GeV.
(c) Cumulative distribution of photon energies.  The dashed
line shows the distribution without lensing; the dotted line
shows the predicted distribution with lensing; while the solid
line shows the distribution for a mock observation of 100
photons.  A Kolmogorov-Smirnov test rejects the hypothesis
that this gamma-ray burst was not lensed at more than 99\%
confidence.
}
\label{fig:GRBspec}
\end{figure}

It is instructive to assess whether GLAST could detect the
fringe pattern in an attolensed gamma-ray burst.  At high
energies, a typical gamma-ray burst has a power law spectrum
such that the number of photons per unit energy is
\beq
  \frac{dN_0}{dE} \propto E^{-\eta}\,,
\eeq
with $\eta \approx 2.25$ \cite{preece}.  If the gamma-ray
burst is attolensed, its observed energy spectrum equals the
intrinsic spectrum multiplied by the energy-dependent lensing
magnification $\mm$ from eq.~(\ref{eq:mm2}),
\beq
  \frac{dN_{\rm obs}}{dE} \propto \mm(E)\,E^{-\eta}\,.
\eeq
\reffig{GRBspec}a shows a sample energy spectrum for a
gamma-ray burst attolensed by a braneworld black hole with
mass $\Mbh = 10^{-18}\,M_\odot$, for a configuration in
which the scaled source position is $\beta = 0.75$.  The
dark fringes (valleys) appear to be more prominent than the
bright fringes (peaks), although this is an artifact of using
a logarithmic vertical scale.  \reffig{GRBspec}b then shows an
example of a binned spectrum as it might be observed by GLAST.
In this example, the bin width is about three times GLAST's
energy resolution, and the errorbars indicate the statistical
uncertainties if there are 100 photons detected with energies
between 30 MeV and 1 GeV.

The ``observed'' spectrum shows an excess near 30--40 MeV,
and a clear dip near 100 MeV, relative to the unlensed spectrum.
To determine the significance of such features, and more
generally to assess the ability to distinguish between the
unlensed and attolensed cases, we use the Kolmogorov-Smirnov
test to compare the ``observed'' spectrum to the unlensed
power law spectrum.  (The KS test applies to the cumulative
distribution of photon energies, shown in \reffig{GRBspec}c.)
For this combination of mass and source position, 100 counts
is typically sufficient to reject the hypothesis that the
gamma-ray burst was not lensed at more than 99\% confidence.
While the number of counts required to reach 99\% confidence
varies with the black hole mass, and to a lesser extent the
source position, it is important to see that for a typical
scenario strong results could be obtained with as few as
100 counts.  GLAST is expected to see of order 20 gamma-ray
bursts per year with more than 100 counts above 30 MeV
\cite{omodei}.

\subsection{On Identifying Primordial Braneworld Black Holes}

Explicit braneworld effects (characterized by $\vel^2$) are
vanishingly small for cosmological scenarios, although they
might be measurable in Solar System scenarios.  This does not
limit our ability to test braneworld gravity, though.  The
reason is that the braneworld model makes very different
predictions, compared with general relativity, about the
formation and survival of primordial black holes
\cite{primordial-1,primordial-2,primordial-3}.  The starkest
difference is that in GR all primordial black holes smaller
than $\sim\!10^{-19}\,M_\odot$ would have evaporated by the
present day \cite{GRevap}, while in braneworld gravity black
holes as small as $\sim 1 \mbox{ kg} \sim 10^{-30}\,M_\odot$
may be able to survive to today \cite{primordial-3,MajMuk1}.
Therefore, a clear detection of a black hole with mass
$\lesssim 10^{-19}\,M_\odot$ would violate the GR prediction
and support the braneworld model.

Observing attolensing fringes would not only reveal a
primordial black hole but also place an important upper limit
on its mass.  From eq.~(\ref{eq:Ebri2}), the energies of the
bright and dark fringes are given by $(1+z_L)^{-1} \Mbh^{-1}$
times a dimensionless factor of order unity that depends on
the source position, $\beta$.  The source position could be
determined from the fringe amplitudes (see eq.~\ref{eq:Mvis}).
Thus, analyzing the fringes would fix the combination
$\hat\Mbh \equiv (1+z_L) \Mbh$.  While it may be difficult or
impossible to determine $z_L$, we would know that $z_L \ge 0$
and hence $\Mbh \le \hat\Mbh$.  Since we would be seeking
evidence that there are black holes below some mass threshold,
upper limits available from attolensing would be useful and
important.

\section{Conclusions}
\label{sec:concl}

We have presented a rigorous and comprehensive analytical
formalism for gravitational lensing due to a braneworld black
hole described by the Garriga-Tanaka metric.  Using invariant
quantities, we calculated all the fundamental geometric
optics lensing observables.  We then used these results to
pursue a new direction in braneworld black hole lensing:
wave optics.  We computed the total magnification in the
limit of semi-classical wave optics, and gave explicit
formulas for the locations of the bright and dark interference
fringes in the energy spectrum of a source lensed by a
braneworld black hole.

Applying our results to realistic examples of lensing indicates
that traditional lensing scenarios involving stellar-mass or
supermassive black holes will be unable to test braneworld
gravity in the foreseeable future.  However, attolensing by
primordial braneworld black holes does provide a powerful
opportunity to probe braneworld effects via interference
fringes in the energy spectra of gamma-ray bursts.  If
primordial braneworld black holes contribute a non-negligible
fraction of the dark matter, there should be many within our
Solar System.  These nearby primordial black holes could be
used to test braneworld gravity directly by looking for the
$\vel^2$ correction terms in the interference pattern, and
indirectly by looking for primordial black holes smaller than
the evaporation limit predicted for general relativity.
Primordial black holes would also be spread throughout the
cosmos.  They may produce a measurable probability for
attolensing of gamma-ray bursts, which would again afford
the possibility of detecting black holes smaller than the GR
limit.

It is worth reiterating that a population of primordial
braneworld black holes in the Solar System is not ruled out by
current dynamical constraints \cite{anderson,gron}.  In the
future, the Laser Interferometry Space Antenna (LISA) may be
able to detect gravitational impulses from passing primordial
black holes, but LISA's sensitivity and noise from the Moon
limit the detectability to $\Mbh \gtrsim 10^{-19}\,M_\odot$
\cite{LISA-PBH}.  Cosmologically, the lack of obvious
femtolensing in a sample of 118 gamma-ray bursts places a
weak upper limit on the cosmological density of primordial
black holes in the mass range $10^{-16}$--$10^{-13}\,M_\odot$
\cite{marani}; but the current constraint neither excludes
fractions as high as $\Obh \sim 0.1$ nor probes masses smaller
than about $10^{-16}\,M_\odot$.  In summary, current
gravitational constraints do not rule out a substantial
population of primordial black holes; for the future,
attolensing may be the only way to probe the important mass
scale below $10^{-19}\,M_\odot$.

While our analysis has been as rigorous and comprehensive as
possible, there is one significant limitation.  Since the full
metric describing the spacetime around a braneworld black hole
is still unknown, we cannot extend our analysis of lensing to 
the physical optics regime at the present time.  We hope that
this new fundamental test of braneworld gravity will motivate
further attempts to determine the full metric for a braneworld
black hole.

\begin{acknowledgments}

We thank Tom Banks, Ernesto Eiroa, Archan Majumdar, Takahiro Tanaka,
and Richard Whisker for helpful correspondences on braneworld
gravity.  We thank the anonymous referee for valuable suggestions.
This work was supported in part by NSF grants DMS-0302812, AST-0434277,
and AST-0433809.
 
\end{acknowledgments}


\begin{thebibliography}{}


\bibitem{SEF}
P. Schneider, J. Ehlers, and E. E. Falco, 
{\em Gravitational Lenses} (Berlin: Springer, 1992). 

\bibitem{PLW}
A. O. Petters, H. Levine, and J. Wambsganss,
{\em Singularity Theory and Gravitational Lensing} 
(Boston: Birkhauser, 2001).

\bibitem{Saas-Fee}
C. S. Kochanek, P. Schneider, and J. Wambsganss,
{\em Gravitational Lensing: Strong, Weak, and Micro.
Lecture Notes of the 33rd Saas-Fee Advanced Course},
ed. G. Meylan, P. Jetzer, and P. North
(Berlin: Springer-Verlag).



\bibitem{paperI}
C. R. Keeton and A. O. Petters,
Phys. Rev. D {\bf 72}, 104006 (2005).

\bibitem{paperII}
C. R. Keeton and A. O. Petters,
Phys. Rev. D {\bf 73}, 044024 (2006).



\bibitem{RS}
L. Randall and R. Sundrum, Phys. Rev. Lett. {\bf 83}, 4690 (1999).

\bibitem{measure-ell}
C. D. Hoyle, U. Schmidt, B. R. Heckel, E. G. Adelberger, J. H. Gundlach,
D. J. Kapner, and H. E. Swanson, Phys. Rev. Lett. {\bf 86}, 1418 (2001);
J. Long, H. Chan, A. Churnside, E. Gulbis, M. Varney, and J. Price,
Nature {\bf 421}, 922 (2003).

\bibitem{bw-lhc}
T. Banks and  W. Fischler, hep-th/9906038 (1999);
R. Emparan, G. T. Horowitz, and R. C. Myers, Phys. Rev. Lett. {\bf 85},
499 (2000);
S. Dimopoulos and G. Landsberg, Phys. Rev. Lett. {\bf 87}, 161602 (2001); 
L. Anchordoqui and H. Goldberg, Phys. Rev. D {\bf 67}, 064010 (2003);
M. Cavaglia, S. Das, and R. Maartens, Class. Quant. Grav. 
{\bf 20}, L205 (2003);
A. Chamblin, F. Cooper, and G. C. Nayak, Phys. Rev. D {\bf 69},
065010 (2004).

\bibitem{bw-showers}
J. L. Feng and A. D. Shapere, Phys. Rev. Lett. {\bf 88}, 021303 (2002);
L. Anchordoqui and H. Goldberg, Phys. Rev. D {\bf 65}, 047502 (2002);
E. J. Ahn, M. Ave, M. Cavaglia, and A. V. Olinto, Phys. Rev. D {\bf 68},
043004 (2003).



\bibitem{PBHspec-1}
S. W. Hawking, Mon. Not. R. Astron. Soc. {\bf 152}, 75 (1971);
B. J. Carr and S. W. Hawking, Mon. Not. R. Astron. Soc. {\bf 168}, 399 (1974);
B. J. Carr, Astrophys. J. {\bf 201}, 1 (1975).

\bibitem{PBHspec-2}
Y. Sendouda, S. Nagataki, and K. Sato, astro-ph/0603509 (2006).



\bibitem{GRevap}
D. N. Page, Phys. Rev. D {\bf 13}, 198 (1976);
D. N. Page, Phys. Rev. D {\bf 14}, 3260 (1976).


\bibitem{primordial-1}
R. Guedens, D. Clancy, and A. R. Liddle, Phys. Rev. D {\bf 66}, 043513 (2002).

\bibitem{primordial-2}
R. Guedens, D. Clancy, and A. R. Liddle, Phys. Rev. D {\bf 66}, 083509 (2002).

\bibitem{primordial-3}
A. S. Majumdar, Phys. Rev. Lett. {\bf 90}, 031303 (2003).



\bibitem{MajMuk1}
A. S. Majumdar and N. Mukherjee, Intl. Jnl. Mod. Phys. D {\bf 14}, 1095 (2005).



\bibitem{GT}
J. Garriga and T. Tanaka, Phys. Rev. Lett. {\bf 84}, 2778 (2000).

\bibitem{GKR}
S. Giddings, E. Katz, and L. Randall, JHEP {\bf 0003}, 023 (2000).

\bibitem{SSM}
M. Sasaki,  T. Shiromizu, and K. I. Maeda, Phys. Rev. D {\bf 62},
024008 (2000).

\bibitem{GWBD}
R. Gregory, R. Whisker, K. Beckwith,  and C. Done,
J. Cosmology and Particle Physics, {\bf 10} (013), 1 (2004).



\bibitem{KS}
S. Kar and M. Sinha, Gen. Relativ. Gravit. {\bf 35}, 1775 (2003).

\bibitem{MajMuk2}
A. S. Majumdar and N. Mukherjee, Mod. Phys. Lett. A {\bf 20}, 2487 (2005).

\bibitem{Eiroa}
E. F. Eiroa, Phys. Rev. D {\bf 71}, 083010 (2005).

\bibitem{Whisker}
R. Whisker, Phys. Rev. D {\bf 71}, 064004 (2005).


\bibitem{virbhadra}
K. S. Virbhadra, D. Narasimha, and S. M. Chitre, Astron. Astrophys.
{\bf 337}, 1 (1998);
K. S. Virbhadra and G. F. R. Ellis, Phys. Rev. D {\bf 62}, 084003 (2002).

\bibitem{petters-SgrA}
A. O. Petters, Mon. Not. R. Astron. Soc. {\bf 338}, 457 (2003).



\bibitem{nakamura}
T. Nakamura and S. Deguchi,
Prog. Theor. Physics. Suppl., {\bf 133}, 137 (1999).



\bibitem{bean}
R. Bean and J. Magueijo, Phys. Rev. D. {\bf 66}, 063505 (2002).

\bibitem{ghez}
A. M. Ghez, S. Salim, S. D. Hornstein, A. Tanner, J. R. Lu, M. Morris,
E. E. Becklin, and G. Duch\^ene, Astrophys. J. {\bf 620}, 744 (2005).

\bibitem{eisen}
F. Eisenhauer, R. Sch\"odel, R. Genzel, T. Ott, M. Tecza, R. Abuter,
A. Eckart, and T. Alexander, Astrophys. J. {\bf 597}, L121 (2003).

\bibitem{micro-pac}
B. Pacy\'nski, Ann. Rev. Astron. Astrophys. {\bf 34}, 419 (1996)

\bibitem{GP}
S. Gaudi and A. O. Petters, Astrophys. J. {\bf 574}, 970 (2002);
S. Gaudi and A. O. Petters, Astrophys. J. {\bf 580}, 468 (2002).

\bibitem{J0737}
M. Burgay et al., Nature {\bf 426}, 531 (2003);
G. Lyne et al., Science {\bf 303}, 1153 (2004).

\bibitem{rafikov}
D. Lai and R. R. Rafikov, Astrophys. J. {\bf 621}, L41 (2005);
R. R. Rafikov and D. Lai, astro-ph/0503461 (2005);
R. R. Rafikov and D. Lai, astro-ph/0512417 (2005).



\bibitem{gould}
A. Gould,
Astrophys. J. {\bf 386}, L5 (1992).

\bibitem{stanek}
K. Z. Stanek, B. Paczy\'nski, and J. Goodman, 
Astrophys. J, {\bf 413}, L7 (1993).

\bibitem{jarosynski}
M. Jarosy\'nski and B. Paczy\'nski,
Astrophys. J. {\bf 455}, 443 (1995).

\bibitem{ulmer}
A. Ulmer and J. Goodman,
Astrophys. J. {\bf 442}, 67 (1995).

\bibitem{gouldgaudi}
A. Gould and B. S. Gaudi,
Astrophys. J. {\bf 486}, 687 (1997).



\bibitem{olling}
R. P. Olling and M. R. Merrifield, Mon. Not. R. Astron. Soc. {\bf 326},
164 (2001).



\bibitem{hilton}
J. L. Hilton, Astron. J. {\bf 117}, 1077 (1999).



\bibitem{anderson}
J. D. Anderson et al., Astrophys. J. {\bf 342}, 539 (1989);
J. D. Anderson et al., Astrophys. J. {\bf 448}, 885 (1995).

\bibitem{gron}
\O. Gr\o n and H. H. Soleng, Astrophys. J. {\bf 456}, 445 (1996).


\bibitem{PressGunn}
W. H. Press and J. E. Gunn,
Astrophys. J. {\bf 185}, 397 (1973);
E. L. Turner, J. P. Ostriker, and J. R. Gott,
Astrophys. J. {\bf 284}, 1 (1984).

\bibitem{holz}
D. E. Holz and R. M. Wald, Phys. Rev. D {\bf 58}, 063501 (1998).


\bibitem{GLAST}
{\tt http://glast.gsfc.nasa.gov}

\bibitem{INTEGRAL}
{\tt http://isdc.unige.ch}

\bibitem{preece}
R. D. Preece et al., Astrophys. J. {\bf 496}, 849 (1998);
R. D. Preece et al., Astrophys. J. Supp. {\bf 126}, 19 (2000).

\bibitem{omodei}
N. Omodei et al., astro-ph/0603762 (2006).


\bibitem{LISA-PBH}
N. Seto \& A. Cooray, Phys. Rev. D {\bf 70}, 063512 (2004).
A. W. Adams \& J. S. Bloom, astro-ph/0405266 (2004).


\bibitem{marani}
G. F. Marani et al., Astrophys. J. Lett. {\bf 512}, 13 (1999).


\end{thebibliography}
\end{document}